\documentclass[aaspp4,11pt,preprint,apjfonts,natbib209]{emulateapj}

\usepackage{graphicx}
\usepackage{times}
 \usepackage{xspace}
\usepackage{amsmath}

\newcommand{\unit}[1]{\mathrm{#1}}
\newcommand*{\Avstel}{\ensuremath{A_{\unit{V\!,\,star}}}\xspace}
\newcommand*{\Avemis}{\ensuremath{A_{\unit{V\!,\,\textsc{Hii}}}}\xspace}
\newcommand*{\Astel}{\ensuremath{A_{\unit{star}}}\xspace}
\newcommand*{\Aemis}{\ensuremath{A_{\unit{\textsc{Hii}}}}\xspace}
\newcommand*{\Av}{\ensuremath{A_{\unit{V}}}\xspace}

\newcommand*{\Avextra}{\ensuremath{A_{\unit{V\!,\,extra}}}\xspace}

\newcommand*{\logtaumin}{\ensuremath{\log_{10}(\tau_{\unit{min}}/\unit{yr})}\xspace}
\newcommand*{\logtau}{\ensuremath{\log_{10}(\tau/\unit{yr})}\xspace}
\newcommand*{\taumin}{\ensuremath{\tau_{\unit{min}}}\xspace}
\newcommand*{\Halpha}{\ensuremath{\unit{H\alpha}}\xspace}
\newcommand*{\Hbeta}{\ensuremath{\unit{H\beta}}\xspace}

\shorttitle{Dust measurements of SF galaxies at \lowercase{$z \sim 1.5$} in 3D-HST}
\shortauthors{Price et al.}
\slugcomment{Accepted for publication in the Astrophysical Journal}

\begin{document}

\title{Direct measurements of dust attenuation in \lowercase{$z \sim 1.5$} 
star-forming galaxies from 3D-HST:\\ 
Implications for dust geometry and star formation rates}

\author{
Sedona H. Price\altaffilmark{1,*}, 
Mariska Kriek\altaffilmark{1}, 
Gabriel B. Brammer\altaffilmark{2}, 
Charlie Conroy\altaffilmark{3}, 
Natascha M. F\"{o}rster Schreiber\altaffilmark{4}, 
Marijn Franx\altaffilmark{5}, 
Mattia Fumagalli\altaffilmark{5}, 
Britt Lundgren\altaffilmark{6}, 
Ivelina Momcheva\altaffilmark{7}, 
Erica J. Nelson\altaffilmark{7}, 
Rosalind E. Skelton\altaffilmark{8}, 
Pieter G. van Dokkum\altaffilmark{7}, 
Katherine E. Whitaker\altaffilmark{9}, 
Stijn Wuyts\altaffilmark{4}
}

\altaffiltext{*}{sedona@berkeley.edu}
\altaffiltext{1}{Astronomy Department, University of California, Berkeley, CA 94720, USA}
\altaffiltext{2}{Space Telescope Science Institute, 3700 San Martin Drive, Baltimore, MD 21218, USA}
\altaffiltext{3}{Department of Astronomy \& Astrophysics, University of California, 
Santa Cruz, CA 95064, USA}
\altaffiltext{4}{Max-Planck-Institut f\"{u}r extraterrestrische Physik, 
Giessenbachstrasse, D-85748 Garching, Germany}
\altaffiltext{5}{Leiden Observatory, Leiden University, P.O. Box 9513, 2300 RA Leiden, 
The Netherlands}
\altaffiltext{6}{Department of Astronomy, University of Wisconsin, 475 N Charter Street, 
Madison, WI 53706, USA}
\altaffiltext{7}{Department of Astronomy, Yale University, New Haven, CT 06511, USA}
\altaffiltext{8}{South African Astronomical Observatory, P.O. Box 9, Observatory 7935, South Africa}
\altaffiltext{9}{Astrophysics Science Division, Goddard Space Flight Center, Code 665, 
Greenbelt, MD 20771, USA}


\begin{abstract}
The nature of dust in distant galaxies is not well understood, and until recently few direct dust 
measurements have been possible. We investigate dust in distant star-forming galaxies 
using near-infrared grism spectra of the 3D-HST survey combined with archival multi-wavelength 
photometry. 
These data allow us to make a direct comparison between dust around star-forming regions 
(\Avemis) and the integrated dust content (\Avstel). We select a sample of 163 galaxies 
between $1.36\le{}z\le1.5$ with \Halpha signal-to-noise ratio $\ge5$ and measure 
Balmer decrements from stacked spectra to calculate \Avemis. 
First, we stack spectra in bins of \Avstel, and find that $\Avemis=1.86\,\Avstel$, 
with a significance of $\sigma=1.7$. 
Our result is consistent with the two-component dust model, in which galaxies 
contain both diffuse and stellar birth cloud dust. 
Next, we stack spectra in bins of specific star formation rate ($\log\,\unit{SSFR}$), 
star formation rate ($\log\,\unit{SFR}$), and stellar mass ($\log{}M_*$). 
We find that on average \Avemis increases with SFR and mass, but decreases with 
increasing SSFR. 
Interestingly, the data hint that the amount of extra attenuation decreases with 
increasing SSFR. 
This trend is expected from the two-component model, as the extra attenuation 
will increase once older stars outside the star-forming regions 
become more dominant in the galaxy spectrum.
Finally, using Balmer decrements we derive dust-corrected \Halpha SFRs, and find 
that stellar population modeling produces incorrect SFRs if rapidly declining star 
formation histories are included in the explored parameter space. 
\end{abstract}

\keywords{dust, attenuation --- galaxies: evolution --- galaxies: high-redshift}

\maketitle

\section{Introduction}

While dust makes up only a very small fraction of the baryonic mass in galaxies \citep{Draine07}, 
it leaves a large signature on their spectral energy distributions (SEDs). 
Dust extinguishes light in a wavelength-dependent way, and therefore 
distorts the intrinsic SED of galaxies. This distortion, or net dust attenuation, 
may depend on the properties of the dust, the dust-to-star geometry, or both quantities. 
Therefore, recovering the intrinsic stellar SEDs from observations requires 
a thorough understanding of both factors. 

 
Dust properties and geometry are studied using observations of both 
dust emission and attenuation.  
Studies of dust emission at mid- and far-infrared wavelengths have placed constraints 
on the composition, distribution, and mass of dust in nearby galaxies 
\citep[e.g.,][]{Draine07, Galliano08, Dale12}. 
Measurements of dust attenuation are also needed to completely characterize the nature 
of dust, including the dust-to-star geometry.

One method of constraining the dust-to-star geometry is by comparing the 
integrated dust attenuation with the attenuation towards star-forming (SF) regions. 
Dust attenuation affecting the stellar continuum, 
$\Astel$, has been measured using a 
number of methods. These include 
\emph{(i)} line of sight measurements (e.g., the MW, SMC, LMC), 
\emph{(ii)} matching attenuated galaxies with unattenuated galaxies with similar 
intrinsic stellar populations \citep[i.e.][]{Calzetti00, Wild11}, 
\emph{(iii)} fitting the SED with stellar population synthesis models, including a 
prescription for dust, and 
\emph{(iv)} the $L_{\unit{IR}}/L_{\unit{UV}}$ ratio (also known as IRX), 
which probes dust attenuation using energy conservation. 
The latter ratio is directly related to the UV continuum slope $\beta$ \citep{Meurer99}, 
which may be used to infer the dust content for galaxies for which no IR data is available.
(See \citealt{Conroy13} for more discussion on this topic). 

Dust attenuation towards SF regions, $\Aemis$, has also been extensively studied in low-redshift 
galaxies. \Aemis is most directly probed with recombination line flux ratios, 
often using the Balmer decrement H$\alpha$/\Hbeta. The intrinsic line ratio can be calculated 
given reasonable environmental parameters. As dust attenuation is wavelength dependent, 
the measured line ratio compared with the intrinsic ratio combined with an assumed dust 
law yields a measure of the amount of dust attenuation towards SF regions.
This method was used by a number of studies to measure attenuation towards 
SF regions in nearby galaxies \citep[e.g.,][]{Calzetti00, Brinchmann04, GB10}.

By comparing \Avemis and \Avstel, \citet{Calzetti00} find that there is 
extra dust attenuation towards star-forming regions relative to the integrated dust 
content for local starburst galaxies. \citet{Wild11} expand on this work by finding that 
the amount of extra attenuation increases with the axial ratio 
and decreases with SSFR. This implies that 
the dust content of galaxies might have two components 
\citep[e.g.,][]{Calzetti94,Charlot00,Granato00}: 
a component associated with the short-lived birth clouds in SF regions and 
a diffuse component distributed throughout the ISM. 
In this model, the diffuse dust attenuates light from all stars, while the 
birth cloud dust component only attenuates light originating from the SF region.

At higher redshifts, the nature of dust attenuation is much more poorly understood. 
Much of the work on dust attenuation in high-$z$ galaxies has focused 
on the UV slope 
\citep[e.g.,][also see \citealt{Shapley11} for a comprehensive review]{Reddy06, Reddy10, 
Wilkins11, Bouwens12, Finkelstein12, Reddy12, Hathi13}, 
as it is relatively easy to observe. 
However, deviations from the \citet{Meurer99} IRX-$\beta$ relation have been found 
for various galaxy samples \citep[e.g.,][]{Kong04,Johnson07,Conroy10,Gonzalez-Perez13}. 
Additional methods of measuring attenuation in high-$z$ galaxies include 
SED modeling of photometric or spectroscopic observations \citep[e.g.,][]{Buat12, Kriek13} 
andcomparison of star formation rate (SFR) indicators \citep[e.g.,][]{Wuyts13}.

Direct measurements of dust attenuation toward \textsc{Hii} regions using Balmer 
decrements are very challenging for $z>0.5$ as both \Halpha and \Hbeta are shifted 
to the less-accessible near-infrared window. Careful survey design and instrument 
improvements have made measurements of the Balmer decrement possible for larger 
samples of intermediate redshift galaxies, e.g., between $0.4 \lesssim z \lesssim 1$ 
\citep{Savaglio05}, $z \sim 0.5$ \citep{Ly12}, and $z \sim 0.8$ \citep{Villar08,Momcheva13}. 
However, until recently Balmer decrements have been measured for only a small number 
of more distant galaxies \citep[e.g.,][]{Teplitz00,vanDokkum05,Hainline09,Yoshikawa10}.

Interestingly, current studies of dust properties in distant star-forming galaxies yield contrasting 
results. \citet{Erb06b} and \citet{Reddy10} find that extra attenuation 
towards \textsc{Hii} regions leads to an overestimate 
of the \Halpha SFR relative to the UV slope SFR. However, \citet{ForsterSchreiber09} 
compare measured and predicted $L_{\Halpha}$ and \Halpha equivalent widths and find the 
best agreement when \Avemis includes extra attenuation relative to \Avstel. A comparison of 
overlapping objects with \citet{Erb06b} shows that the previous aperture corrections might be 
overestimated, which could have masked some extra attenuation. \citet{Yoshikawa10} compare 
\Avstel from SED fitting and \Avemis from Balmer decrements for a small sample and find that 
the high-$z$ objects are consistent with the local universe \citet{Calzetti00} prescription 
for extra dust attenuation. 
Additionally, \citeauthor{Wuyts11a} \citeyearpar{Wuyts11a,Wuyts13}, \citet{Mancini11} 
and \citet{Kashino13} find the best agreement between \Halpha SFRs and UV+IR/SED SFRs 
when extra attenuation is adopted, either the same as the \citet{Calzetti00} relation 
\citep{Wuyts11a, Mancini11} or a slightly lower ratio \citep{Wuyts13, Kashino13}.  
\citet{Kashino13} also measure the Balmer decrement and find 
the amount of extra attenuation is lower than the \citet{Calzetti00} relation.

These contrasting results may not be surprising, given the different and indirect methods 
and/or small and biased samples of most studies. 
Direct measurements of a statistical sample of distant galaxies are required to clarify these dust 
properties. This is now possible, as new NIR instruments with multiple object spectroscopy capabilities are able to measure the Balmer decrement for larger and more complete samples of 
high-redshift objects. In particular, the \emph{Hubble Space Telescope}'s WFC3/G141 grism filter 
provides slit-less spectra, allowing for a non-targeted survey of a large number of high-redshift 
galaxies. The \emph{HST} grisms also avoid atmospheric near-IR absorption.

A number of surveys, including the 3D-HST survey \citep{vanDokkum11,Brammer12}, 
CANDELS \citep{Koekemoer11}, and the WISP survey \citep{Atek10}, have taken advantage of the 
\emph{HST} grism capabilities to survey high redshift galaxies. \citet{Dominguez13} were the first 
to use WFC3 grism data to make measurements of the Balmer decrement on a large, non-targeted 
sample. However, as their sample was not drawn from regions of the sky with existing deep 
photometric coverage, they were unable to examine trends of dust versus integrated galaxy properties.

We present a statistical study of dust attenuation measured using the Balmer decrement 
for a large, non-targeted sample of galaxies at $z \sim 1.5$. 
Both rest-frame optical spectra and deep photometry are available, allowing us to compare 
attenuation towards \textsc{Hii} regions 
with the total integrated dust attenuation and other galaxy properties, 
including stellar mass and SFR. 

Throughout this paper we adopt a $\Lambda$CDM universe with $\Omega_m = 0.3$, 
$\Omega_{\Lambda} = 0.7$, and $H_0 = 70 \ \unit{km \, s^{-1} \, Mpc^{-1}}$.

\begin{figure*}
	\begin{center}
	
	\includegraphics[width=0.95\textwidth]{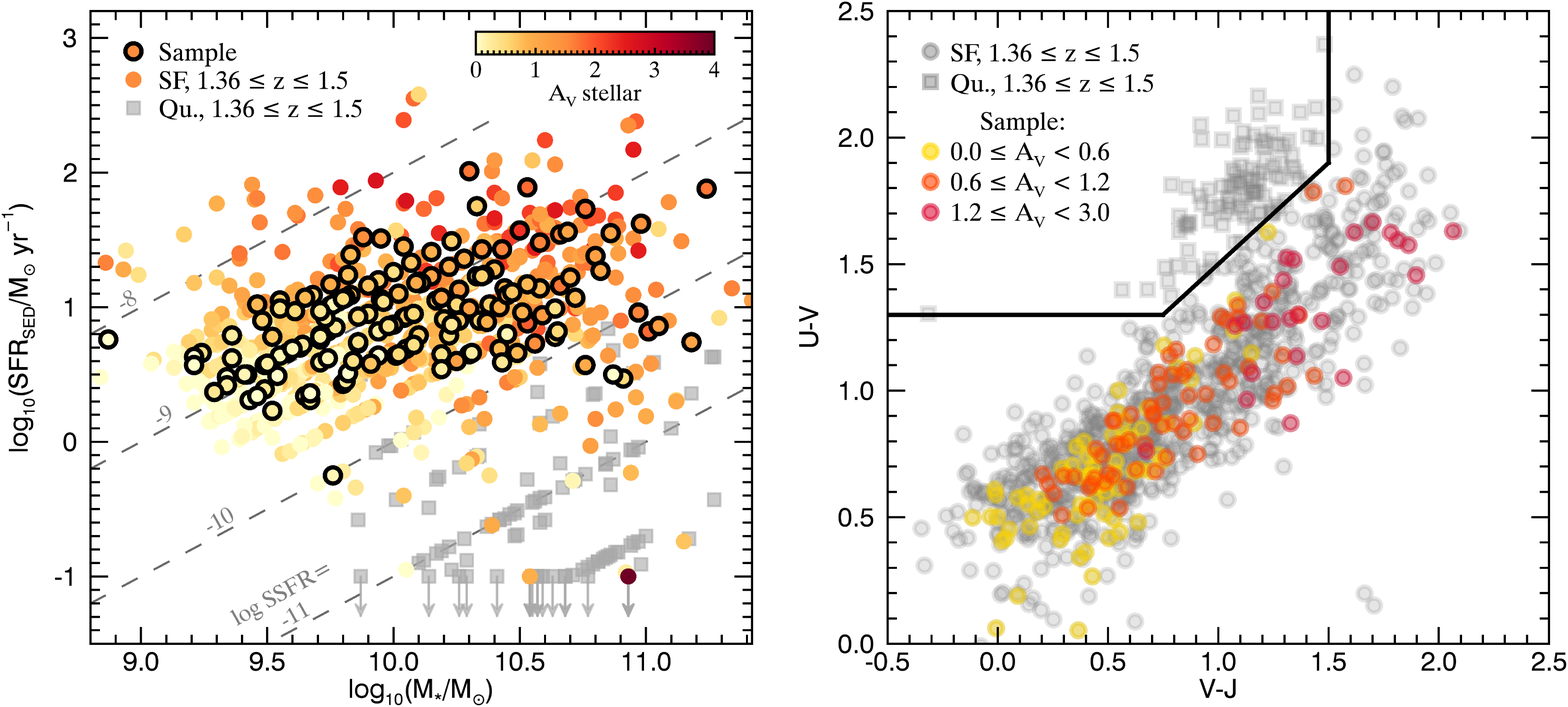}
	\caption{Sample characteristics relative to all 3D-HST galaxies in the same redshift range. 
	The left panel shows $\log M_*$ vs. log SFR (from SED fitting) 	for all 3D-HST 
	galaxies at $1.36 \leq z \leq 1.5$. The black circles indicate the sample selected for 
	direct Balmer decrement measurements, consisting of galaxies with a strong \Halpha 
	detection (i.e. S/N $\geq$ 5). Colors indicate the stellar \Av 
	of the best-fit SED models found using the grism spectra and the photometry, as described 
	in Section \ref{sec:observations}. The grey dashed lines show constant values of log SSFR. 
	The right panel presents the rest-frame U-V and V-J colors for all 3D-HST galaxies between 
	$1.36 \leq z \leq 1.5$ (in grey) and our sample (color-coded by stellar \Av bin). 
	We discard any galaxies lying within 	the quiescent box (using the definition from 
	\protect\citealt{Whitaker12}) 
	from our sample, as the emission lines for these galaxies likely originate from AGNs. 
	In both panels, objects within the quiescent box of the UVJ diagram are shown as grey 
	squares, while star-forming objects are shown as circles.
	}
	\label{fig:samp_char}
	
	\end{center}
\end{figure*}

\section{Data}
\label{sec:data}

\subsection{Observations and catalog}
\label{sec:observations}

Our sample is drawn from the 3D-HST survey \citep{Brammer12}, 
a \emph{Hubble Space Telescope (HST)} Treasury program adding 
ACS and WFC3 slit-less grism observations to the well-covered 
CANDELS \citep{Koekemoer11} fields: 
AEGIS, COSMOS, GOODS-S, and UDS. The 3D-HST data also include 
observations of the GOODS-N field from program GO-11600 (PI: B. Weiner).

In this work we use the 3D-HST WFC3/G141 grism spectra, which cover 
$1.1\unit{\mu m} < \lambda < 1.65 \unit{\mu m}$. 
The raw grism dispersion is $46.5 \, \unit{\AA \, pixel^{-1}}$, but interlacing during data 
reduction improves the dispersion to $\sim \! 23 \, \unit{\AA \,  pixel^{-1}}$ 
(corresponding to about 10 restframe $\unit{\AA \, pixel^{-1}}$ for 
$z \sim 1.5$). 
The G141 grism has a maximum resolution of $R \sim 130$, corresponding to 
$\sim 110 \, \unit{\AA}$ in the middle of our wavelength range.

The 3D-HST survey makes use of existing deep photometric coverage in each of the 
survey fields, combining the grism spectra with multi-wavelength photometric 
data. The 3D-HST photometric catalogs are discussed in detail in \citet{Skelton14}. 
This work uses version 2.1 of the photometric and grism catalogs.

A modified version of the \textsc{Eazy} code \citep{Brammer08} is used on the 
combined grism spectra + photometry to measure the redshifts, emission line fluxes, 
and rest-frame U, V, J fluxes of the individual 3D-HST galaxies. 
Stellar masses, integrated dust attenuation, SFRs, and 
specific star formation rates (SSFRs)  
are determined by fitting stellar population synthesis models to the photometric data 
using the FAST code \citep{Kriek09}.  
We use a separate set of parameters than those used by \citet{Brammer12}, 
for reasons discussed in Section \ref{sec:SFRHa_SFRSED}. 
We use the \citet{bc03} stellar population synthesis models, assuming a \citet{Chabrier03} 
stellar initial mass function, solar metallicity, an exponentially declining star formation history 
with a minimum \textit{e}-folding time of $\logtaumin = 8.5$, 
a minimum age of 40 Myr, and an integrated dust attenuation \Av between 0 and 4 assuming 
the dust attenuation law by \citet{Calzetti00}.

\citet{Brammer12} provide complete details on the 3D-HST survey data reduction 
and parameter measurement procedure.

\subsection{Sample selection}
\label{sec:sample_selection}

We select galaxies in the redshift range $1.36 \leq z \leq 1.5$, for which \Halpha and 
\Hbeta are generously covered by the G141 grism. In addition we impose a signal-to-noise 
ratio (SNR) cut for \Halpha of SNR $\geq 5$, to measure a decent line signal. 
We make no \Hbeta SNR cut, to avoid biasing our sample against the dustiest galaxies.

We have a number of additional selection criteria, to ensure high quality of the spectra. 
First, to avoid cases of line misidentification, the photometric-only and grism+photometry 
redshifts (hereafter referred to as the grism redshifts) must have good 
agreement, i.e. $| z_{\unit{phot}} - z_{\unit{grism}} | \leq 0.2$. 
Second, the contamination from other sources (which is an issue because of the 
slit-less nature of the grism spectra) may not exceed 15\%.
Third, there must be grism coverage of at least 95\% and 
must include the \Halpha and \Hbeta lines. 
Fourth, no more than 50\% of the spectrum may be flagged as problematic (due to bad 
pixels or cosmic rays) during reduction.

To study dust attenuation towards star forming regions, we do not want AGN to contaminate our 
emission lines. To reject AGN, we exclude any objects that have a detected X-ray luminosity 
$L_X > 10^{42} \, \unit{erg \; s^{-1}}$ \citep{Mendez13, Rosario13, Bauer02}
by matching against the Chandra Deep Field 
North and South surveys \citep{Alexander03, Xue11} and the XMM-Newton serendipitous survey 
\citep{Watson09}. Furthermore we reject all objects that fall within the \citet{Donley12} IRAC
AGN region, or that fall within the quiescent box in the UVJ diagram defined in 
\citet{Whitaker12} (as the line emission likely originates from an AGN).

Finally, we visually inspect the grism spectra and photometry of the preliminary sample to 
reject problematic objects 
(i.e. objects with incorrect line identification or poor quality broadband photometry). 
Our final sample includes 163 galaxies. 
Figure \ref{fig:samp_char} shows how our sample compares to the full galaxy distribution 
at a similar redshift. The selected galaxies all have relatively high SFRs, and lie 
along the ``star-forming main sequence'' 
\citep[e.g.,][]{Noeske07, Daddi07, Wuyts11b, Whitaker12b, Nelson13}.

\begin{figure*}
	\begin{center}

	\includegraphics[width=0.92\textwidth]{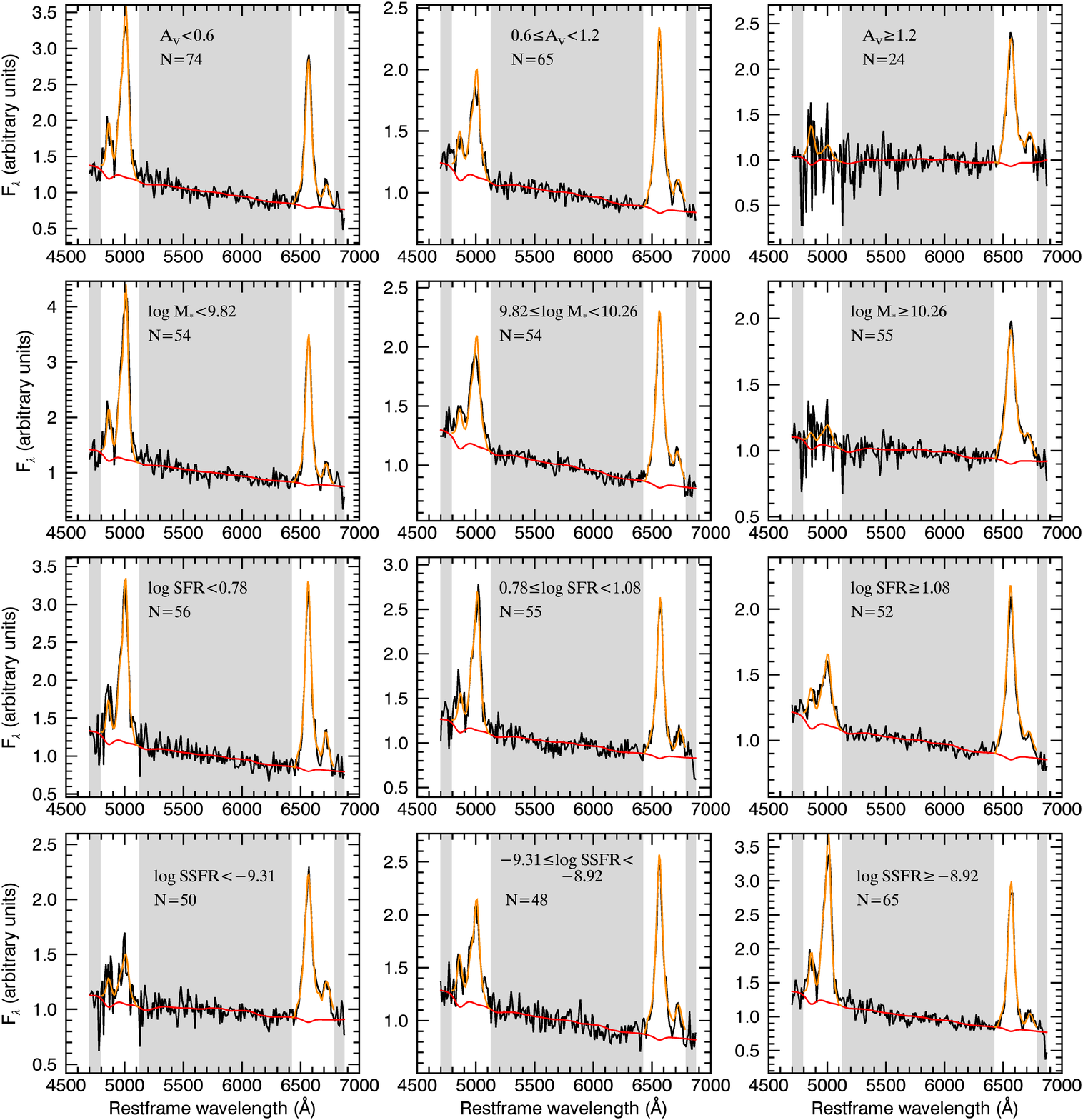}
	\caption{Stacked spectra for bins in stellar $\Av\!$ (top), 
	$\log M_{*}$ (second row), 
	log SFR (third row) and  
	log SSFR (bottom). 
	In each panel, the stacked spectrum is plotted in black, with 
	the continuum fit shown in red. The best-fit line measurements 
	(from shortest to longest wavelength: \Hbeta, [OIII]$\lambda\lambda 4959,5007 \unit{\AA}$, 
	blended H$\alpha$+[NII]$\lambda\lambda 6548,6584 \unit{\AA}$, and [SII]$\lambda\lambda 
	6717,6731 \unit{\AA}$) are shown in orange. 
	The additional continuum correction is done by fitting the portions 
	of the spectra within the shaded grey regions.}
	\label{fig:spec}
	
	\end{center}
\end{figure*}

\subsection{Stacking}
\label{sec:stacking}

The spectra of individual galaxies in this sample are too noisy to yield a clear measurement of 
the Balmer decrement, the ratio of the flux of \Halpha to \Hbeta ($F_{\Halpha}/F_{\Hbeta}$).
Thus, we bin galaxies by parameter (SED \Av, SSFR, SFR, stellar mass) and 
stack the spectra before measuring line fluxes. 

Prior to stacking, we scale the spectra to match the photometry, 
as the spectra and photometry have a slightly different slope for most galaxies. 
A linear correction for this effect is calculated during 
the reduction stage, and we apply this linear correction to the individual grism spectra. 
Individual spectra are also corrected for contamination from other sources 
during the reduction process. 

We adopt a uniform methodology for stacking spectra within a bin. 
First, we only use the portion of the grism spectra that falls between 
$1.13 \unit{\mu m}$ and $1.65 \unit{\mu m}$ (observed wavelength) to avoid noise at the 
edge of the grism coverage. Then the individual spectra are continuum normalized by scaling 
the biweighted mean value of the flux between $5500-6000 \unit{\AA{}}$ (rest-frame) to unity. 
The individual spectra are then interpolated onto a common rest-frame wavelength grid.

The normalized, interpolated spectra are stacked at each wavelength. 
For each bin, simulated spectra ($N = 500$) are generated by perturbing the 
individual spectra, assuming that the flux errors are normally distributed, and then stacking 
the perturbed spectra using the same procedure as above. The simulated spectra are 
used in determining the errors of the emission lines (see Section \ref{sec:line_measurement}).

The best-fit stellar population models are sampled over the same wavelength regime as 
the original spectra and used as the continua.  The best fits are stacked in the same way: 
they are continuum normalized and interpolated onto a common rest-frame wavelength grid, 
then averaged together. We then convolve the stacked continua models with the 
stacked line profiles (discussed in Section \ref{sec:line_measurement}) to match the 
resolution of the grism spectra.

To estimate the error in the continua, we perturb the photometry of an object 
and determine the FAST best-fit model for each perturbation. This procedure is repeated 
a number of times for each object. Within each bin, we randomly select a best-fit model 
to the perturbed photometry for each object and stack to construct a simulated  continuum. 
This procedure is repeated 500 times, in order to create a continuum model for each 
simulated spectrum, as discussed above. The average photometry and stacked best-fit 
models with errors are shown in the Appendix.

Finally, we apply a second-order polynomial correction between the stacked grism 
spectra and the stacked best-fit stellar population continua, to further correct for possible 
mismatches. This small change corrects for otherwise uncharacterized differences across 
the grism spectra. The corrected, stacked spectra and continua for the different bins are 
shown in Figure \ref{fig:spec}.

The adopted continuum normalization scheme leads objects with higher scaled 
$\Halpha$+[NII] (see Section \ref{sec:line_measurement}) fluxes 
to have more weight in our line stack. 
To correctly compare the parameters from SED fitting (\Avstel, stellar mass, SFR, SSFR) with 
values calculated from the stacked lines, we compute the weighted average of each parameter. 
We use the scaled $\Halpha$+[NII] fluxes as the 
individual objects' weights. Errors on the average parameters are estimated with bootstrapping.

\subsection{Line measurement}
\label{sec:line_measurement}

To measure line fluxes for the stacked spectra, we subtract the convolved, stacked continua 
from the tilt-corrected, stacked spectra. 
We then fit the emission lines in the resulting spectra using least-squares minimization.
For our sample's redshift range, 
the grism spectra have rest-frame coverage of the following spectral lines: 
\Hbeta, [OIII]$\lambda\lambda 4959,5007 \unit{\AA}$, \Halpha, 
[NII]$\lambda\lambda 6548,6584 \unit{\AA}$, and 
[SII]$\lambda\lambda 6717,6731 \unit{\AA}$. However, the resolution of the grism data is 
insufficient to separate the \Halpha and [NII] lines, 
so we measure the blended H$\alpha$+[NII]$\lambda\lambda 6548,6584\unit{\AA}$ line.

The grism line shapes are not well described by gaussian profiles. Because the spatial 
resolution of the WFC3 detector ($\sim \!\! 0'' \!\!.13 \, \unit{pixel^{-1}}$) is much greater than the 
spectral resolution, the spectral line profiles are dominated by the object shapes. 
Thus, for each object the line profile is measured by summing the direct image from 
F140W or F160W over the spatial direction, which is perpendicular to the dispersion direction.
The composite line profiles are created by flux-normalizing the individual profiles, 
multiplying each profile by the object's scaled $\Halpha$+[NII] flux (described in Section 
\ref{sec:stacking}), and finally averaging. This method 
yields a composite profile with the same effective weighting of the objects as results 
from the spectrum stacking method. 
Finally, we scale the profile width to match the width of $\Halpha$+[NII], 
yielding the line profile template for each stack. 
The grism spectra have roughly constant spectral resolution. Thus for each line, 
we scale the line profile width by $\lambda_{\unit{line}}/\lambda_{\unit{H\alpha}}$.

Lines in a spectrum may not have the same profile, possibly 
due to dust, age gradients, or AGN contribution \citep[e.g.,][]{Wuyts12}. However, 
the line fits obtained while using the same line profile (with appropriate width and amplitude scaling) 
match the data very well, suggesting that 
assuming a single profile for a stack is a reasonable approximation. 

Because of the low spectral resolution of the grism spectra, we simultaneously fit the [OIII] doublet 
and \Hbeta, and similarly the blended H$\alpha$+[NII] line and the [SII] doublet. We fix the line 
ratio between [OIII]$\lambda 5007 \unit{\AA}$ and [OIII]$\lambda 4959 \unit{\AA}$ to 3:1 to reduce 
the number of degrees of freedom in our fit, and we fix the redshift of all lines to the value 
measured for \Halpha.

We compute the emission line fluxes and ratios from the best-fit line profile parameters. 
The errors on the line fluxes and ratios are estimated using the simulated 
stacked spectra and continua. For each simulation we measure the best-fit line fluxes and 
ratios in a similar fashion as for the real stacked spectrum. The errors on the 
line fluxes and ratios are calculated from the resulting distributions.
The best-fit line measurements for our stacks are shown in Figure \ref{fig:spec}.

\subsubsection{[NII] correction}
\label{sec:nii_correction}

To measure the Balmer decrement, we need to correct the blended H$\alpha$+[NII] line 
for the [NII] contribution. We use the stellar mass versus [NII]$\lambda 6584$/\Halpha 
relation measured in \citet{Erb06a} for galaxies at $z \sim 2$, as our sample covers a 
similar range of masses and SFRs, and is close in redshift.  

In \citet{Erb06a} the stellar masses are calculated using the integrated 
SFH, while we use the current stellar mass. For the galaxies in our sample, which are all 
reasonably young, the mass from the integrated SFH is about 10\% higher than the current stellar 
mass. We use this estimate to scale down the masses given by \citeauthor{Erb06a} to match our 
stellar mass definition.

We interpolate the values \citeauthor{Erb06a} report in Table 2 to estimate the ratio of 
[NII]$\lambda 6584$/\Halpha given the weighted average stellar mass in each bin. 
We assume an intrinsic line ratio of 3:1 between [NII]$\lambda 6584$ and [NII]$\lambda 6548$ 
to scale this ratio to include the second [NII] line. We use the resulting 
ratio to calculate the \Halpha flux in each stack. 
We do not include the systematic errors in the [NII]/\Halpha ratio in our flux errors.

\section{Dust attenuation compared with galaxy properties}
\label{sec:dust_attenuation_galaxy_properties}

\subsection{Measuring dust attenuation towards star-forming regions}
\label{sec:measuring_dust_attenuation}

The Balmer decrement, $\unit{H\alpha/H\beta}$, lets us determine the amount of dust attenuation 
towards star-forming regions by comparing the measured ratio with the expected  
line ratio given the physical conditions of the region. 
We assume that the \textsc{Hii} region has a temperature $T=10^4 \ \unit{K}$, an electron density 
of $n_e = 10^2 \ \unit{cm^{-3}}$, and that the ions undergo case B recombination. 
These assumptions result in an intrinsic ratio of 
$(\unit{H\alpha/H\beta})_{\unit{int}} = 2.86$ \citep{Osterbrock06}. 
We assume the reddening curve $k(\lambda)$ of \citet{Calzetti00}, which gives us 
\begin{equation}
E(\unit{B-V}) = 1.97 \log_{10} \left[ \frac{(\unit{H\alpha/H\beta})_{\unit{obs}}}{2.86} \right].
\end{equation} 
We combine this expression with $R_\unit{V} \equiv \Av/E(\unit{B-V})$, assuming the value 
$R_{\unit{V}} = (4.05 \pm 0.80)$ from \citet{Calzetti00} 
to calculate the attenuation \Avemis from the \Halpha and \Hbeta flux measured 
for each stack.

\begin{figure}
	\begin{center}
	
	\includegraphics[width=0.48\textwidth]{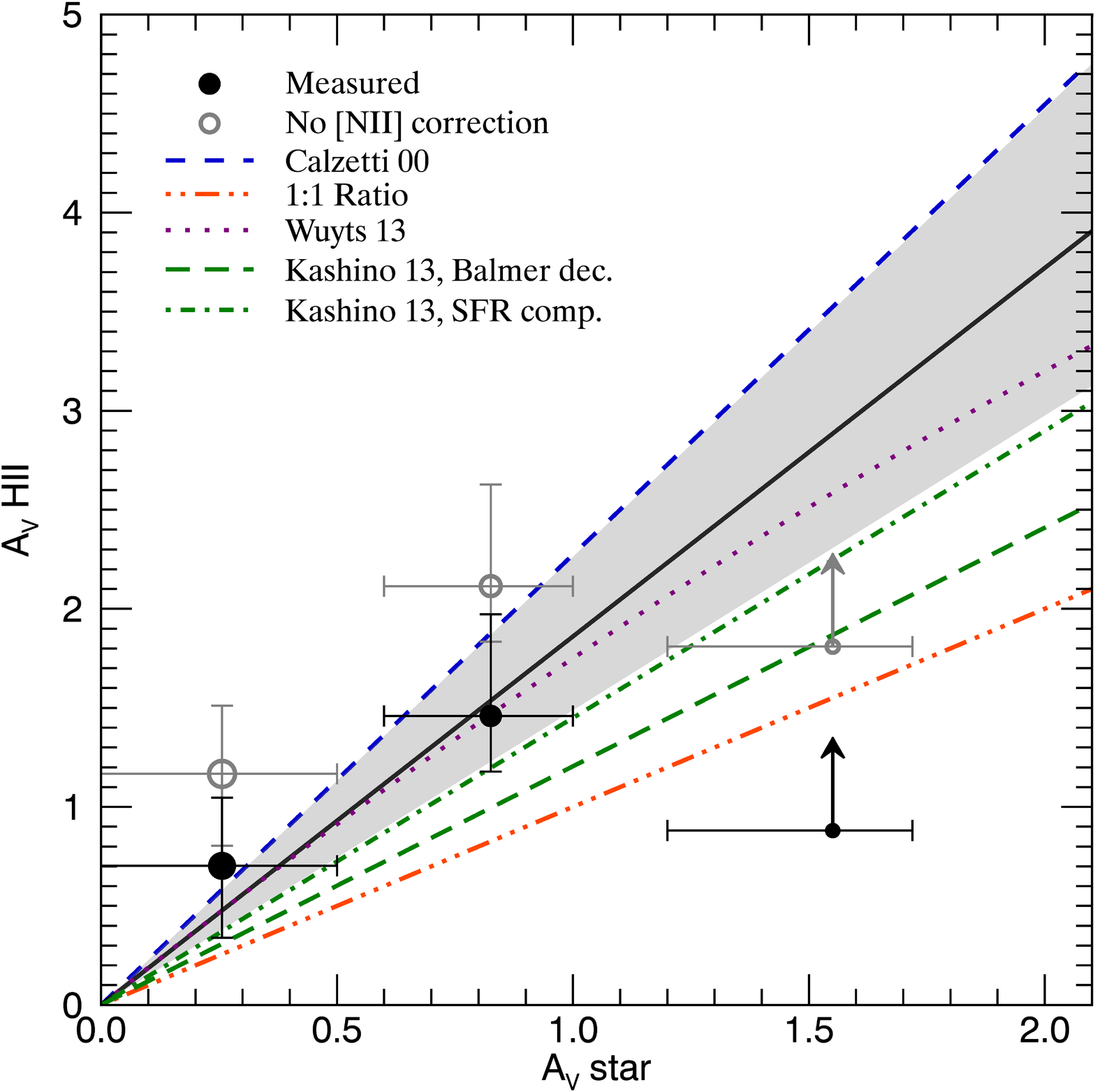}
	
	\caption{Weighted average of individual \Avstel vs. \Avemis measured 
	from the spectra stacked in bins of \Avstel, shown in the top panel of Fig \ref{fig:spec}, 
	both corrected (black circles) and uncorrected (open grey circles) for [NII] contamination. 
	The size of the data points corresponds to the fraction of the 
	scaled $\Halpha$+[NII] flux in each bin. 
	Bins without a significant detection ($\ge 2 \sigma$) of \Hbeta 
	are shown as $2 \sigma$ lower limits. 
	The \Avstel errors shown are the $1\sigma$ scatter within the bins. 
	The black line shows the best-fit line to our [NII] corrected values, 
	which has a slope of 1.86. 
	The fit error is shown with the shaded grey region. 
	The best-fit indicates there is extra attenuation towards emission line regions. 
	The data are consistent with the ratio of \Avemis to \Avstel from \citet{Calzetti00} 
	(blue dashed line), \citet{Wuyts13} (purple dotted line), and 
	both relations found by \citet{Kashino13} (green long dash, dash-dot lines). 
	The data are inconsistent with the assumption of no extra dust attenuation 
	towards emission line regions (orange dash-dot-dot line). 
	}
	\label{fig:avplot}

	\end{center}
\end{figure}

\begin{figure*}
	\begin{center}

	\includegraphics[width=0.95\textwidth]{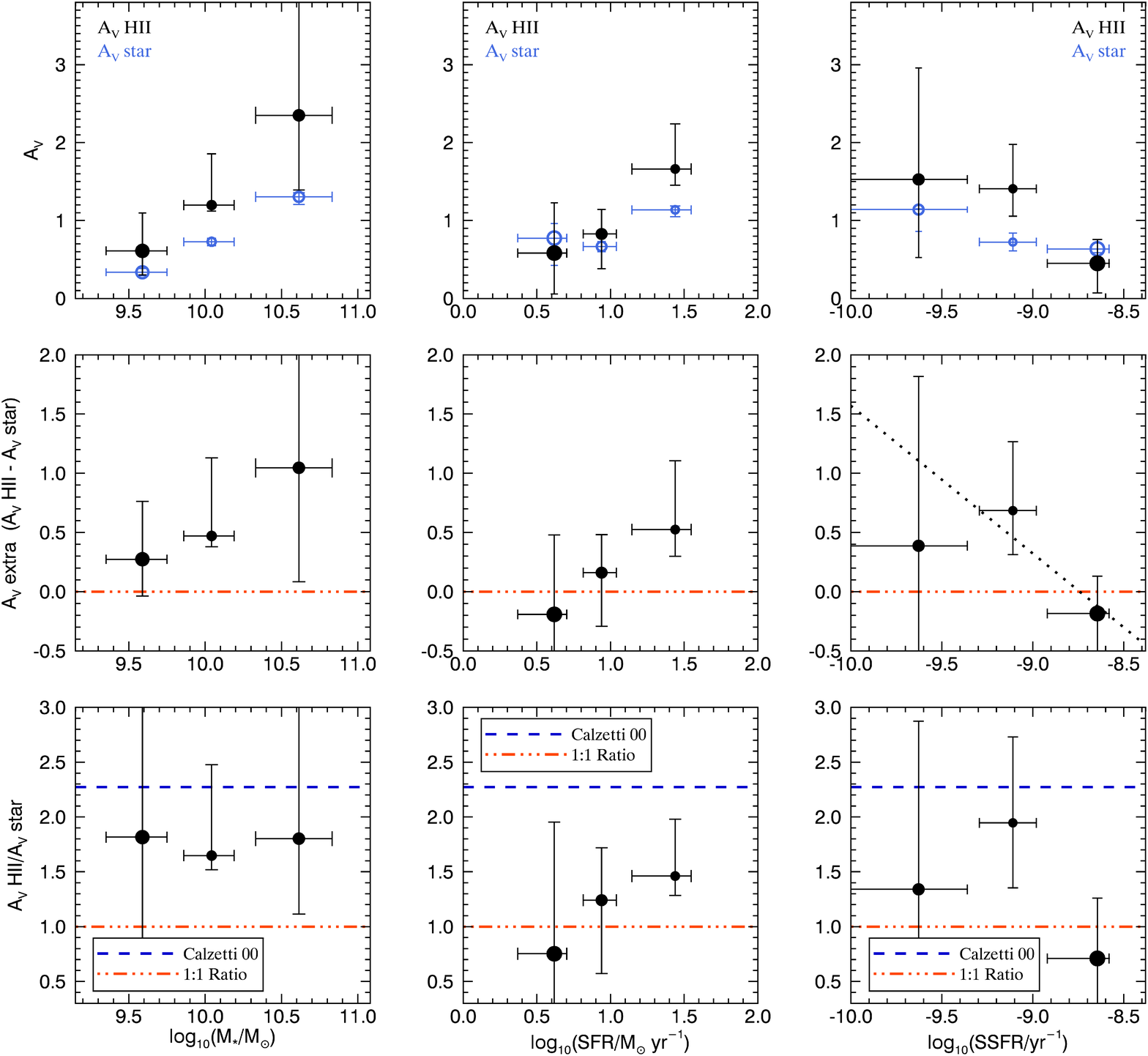}
	\caption{ 
	\Av measurements for bins in $\log M_{*}$, log SFR and log SSFR.
	The top row shows \Avstel and \Avemis vs. the binned parameters.
	The middle row shows the difference between emission and stellar \Av 
	($\Avextra$) vs. the binned parameters. The orange dash-dot-dot line shows the case 
	of no extra attenuation towards star-forming regions. 
	The bottom row shows the ratio of \Avemis to \Avstel 
	vs. the binned parameters. The blue dashed and orange dash-dot-dot lines in 
	the third row are the ratio between emission \Av and stellar \Av used in 
	\protect\citet{Calzetti00} and the case of no extra attenuation towards star-forming 
	regions, respectively. 
	The size of the data points is described in Fig. \ref{fig:avplot}. 
	The errors in $\log M_{*}$, log SFR and log SSFR are the $1\sigma$ scatter within the bins. 
	In the right panel the black dotted line shows the best-fit relation of 
	$\Avextra = \Avemis-\Avstel$ vs. log SSFR, with the fit errors shown with the 
	shaded grey region.
	}
	\label{fig:avratio}
	
	\end{center}
\end{figure*}

\subsection{Integrated stellar \Av}
\label{sec:av_sed}

We first investigate \Avemis in bins of \Avstel, to better constrain the currently contested 
relationship between the integrated dust content and the dust associated with SF regions for 
high-redshift galaxies. We choose bins of \Avstel to probe the full range of integrated stellar 
dust attenuation in our sample, from low to medium to high attenuation. 
We stack the spectra in these bins and measure \Avemis on the stack using the 
relations given in Section \ref{sec:measuring_dust_attenuation}. The results are shown 
in Figure \ref{fig:avplot}. 

We perform a least-squares ratio fit to the data.  
The best-fit relation, assuming $R_\unit{V}$ is the same for the stellar continuum 
and the \textsc{Hii} regions, is 
\begin{equation}
\Avemis = 1.86_{-0.37}^{+0.40} \  \Avstel ,
\end{equation}
indicating that on average \Avemis is 1.86 times higher 
than \Avstel in star-forming galaxies at $z \sim 1.5$. 
This is a slightly lower amount of extra attenuation than the ratio of 2.27 which \citet{Calzetti00} 
find for low redshift star-forming galaxies, but our data are not statistically different 
from the \citet{Calzetti00} relation. 
The data are inconsistent with the assumption of no extra dust attenuation towards 
star-forming regions and are inconsistent with a constant value at the 
$\sigma = 1.2, \ 1.7$ levels, respectively. 

The data are consistent with the results by \citet{Wuyts13}, who find an average relation 
between \Avstel and $\Avextra = \Avemis - \Avstel$ for galaxies at $0.7 \leq z \leq 1.5$. 
Their relation (shown in Figure \ref{fig:avplot} by the dotted purple line) is the dust attenuation 
required for agreement between \Halpha SFRs and UV+IR SFRs, 
or if there was no IR detection, SED SFRs. 
Our data are also consistent with the relations \citet{Kashino13} find by comparing 
UV and \Halpha SFR indicators (green dash-dot line) 
and measured from the Balmer decrement (green long dash line).

We also show the results without correcting the \Halpha flux for [NII] (open grey circles). 
These points overestimate the amount of \Avemis relative to \Avstel, 
demonstrating the necessity of correcting for [NII] when measuring the attenuation of the 
star-forming regions using grism data.

However, it is important to note that there may be considerable 
scatter in the \Avstel and \Avemis values for individual galaxies, 
so this result only holds on average for a collection of galaxies.

\subsection{Stellar mass, SSFR, and SFR}
\label{sec:ssfrsfrmass}

In this section we probe the change in dust properties over bins of stellar mass, SSFR, and 
SFR. We select bin boundaries for each of these properties to 
distribute the number of galaxies as equally as possible 
between the bins.\footnote[10]{The SED parameters are sampled from a discrete array, 
which sometimes results in a large number of objects with the same parameter value.}

The left panels of Figure \ref{fig:avratio} show the results for the stacks in stellar mass, 
including the measured \Avemis and weighted average \Avstel (top)
and the difference between \Avemis and \Avstel (\Avextra, middle) as a function of stellar mass. 
We also plot the ratio $\Avemis/\Avstel$ (bottom) to facilitate comparison with previous 
studies.

We see an increase in both \Avstel and \Avemis with increasing stellar mass,  
inconsistent with a constant value at the $\sigma = 12, \, 1.2$ levels, respectively. 
The plot of \Avextra shows a roughly constant amount of extra attenuation with mass. 
The data are consistent with a constant value of \Avextra. 
The ratio $\Avemis/\Avstel$ is consistent with the value found by \citet{Calzetti00} for all mass 
bins.

The results for stacks in SFR are shown in the center panel of Figure \ref{fig:avratio}. 
Both \Avstel and \Avemis show an increase with SFR, at the $\sigma = 4.3, \, 1.0$ levels, 
respectively. \Avextra is consistent with no trend with SFR. 
As before, we show the ratio $\Avemis/\Avstel$ for direct comparison with past work. 
On average, the ratio is below the value found in \citet{Calzetti00}.

The results for stacks in SSFR are shown in the right panel of Figure \ref{fig:avratio}. 
Both \Avstel and \Avemis decrease with increasing SSFR, and are inconsistent 
with a constant value at the $\sigma = 3.2, \, 1.2$ levels, respectively.

We also find a slight decreasing trend in \Avextra with SSFR, however this trend is not 
significant as the difference between the data and a constant value is only $\sigma = 0.7$. 
Again, we show the ratio $\Avemis/\Avstel$ for comparison. Our ratio is consistent with the 
\citet{Calzetti00} value for the lowest two SSFR bins, while our ratio for the highest SSFR bin 
is lower.

Interestingly, \Avextra is most strongly correlated with SSFR, rather than stellar mass or SFR. 
We perform a least-squares linear fit to \Avextra vs. log SSFR, using an offset in log SSFR to 
avoid correlated intercept and slope errors. We find a best-fit relation of
\begin{equation}
\Avextra = 0.48_{-0.32}^{+0.41} - 1.25_{-0.91}^{+0.87} \,  
\left[\log_{10}(\unit{SSFR/yr^{-1}}) + 9.13 \right].
\end{equation}
This possible trend could be explained by the two-component dust model, 
as we discuss in Section \ref{sec:interpretation}.

\section{Discussion}
\label{sec:discussion}

\subsection{Physical interpretation}
\label{sec:interpretation}

The observed extra attenuation towards emission-line regions, and the decrease 
in the amount of extra attenuation with increasing SSFR, are consistent with a 
two-component dust model \citep[e.g.,][]{Calzetti94,Charlot00,Granato00,Wild11}.
This model assumes there is a diffuse (but possibly clumpy) dust component in the ISM that 
affects both the older stellar populations and star-forming regions, as well as 
a component associated with the short-lived stellar birth clouds that only 
affects the stars within those regions. 

\begin{figure}
	\begin{center}

	\includegraphics[height=0.48\textwidth,angle=-90]{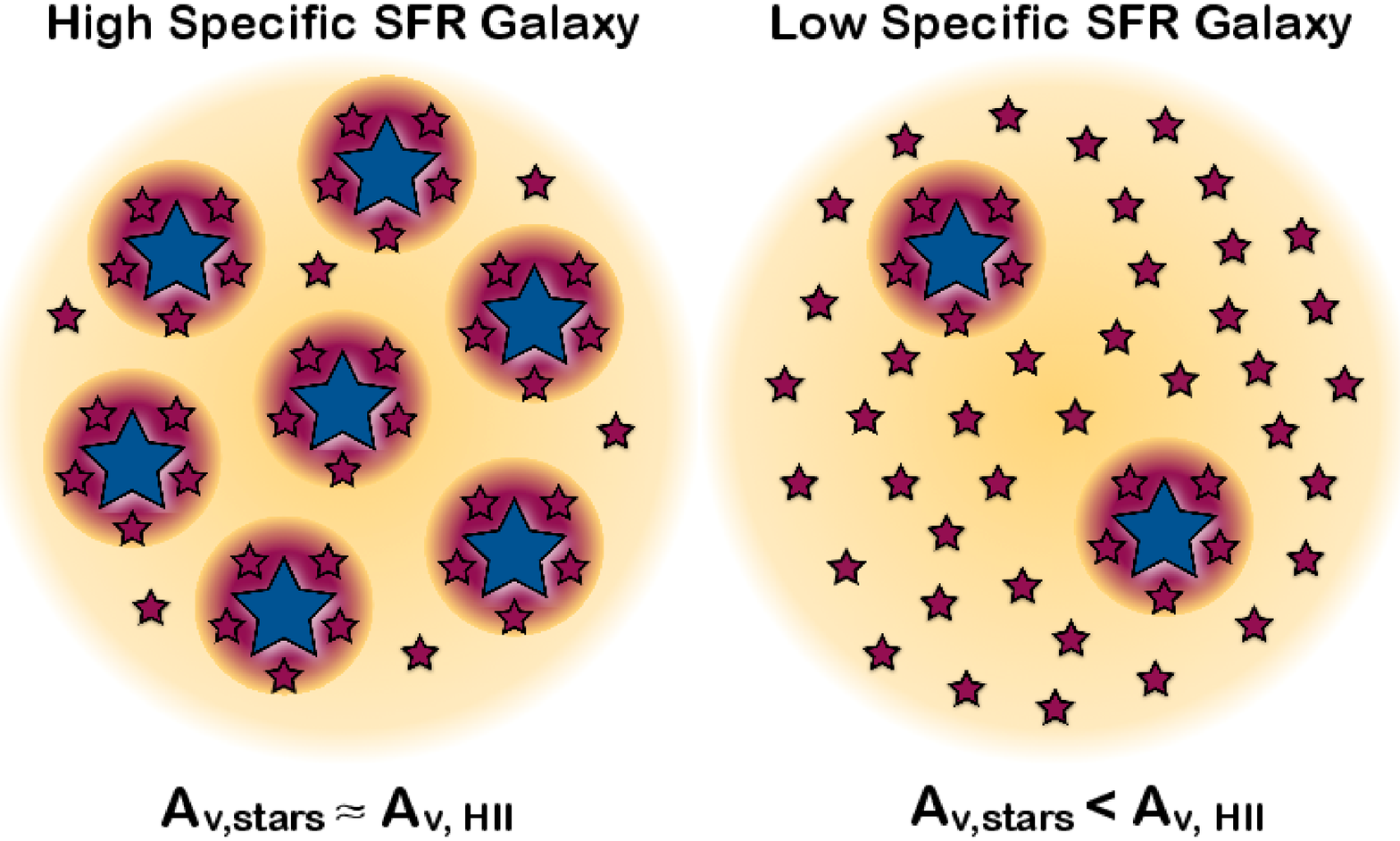}
	\caption{ 
	Illustration of the two-component dust model in galaxies with 
	high (left panel) and low (right panel) specific SFRs. 
	The yellow regions indicate the diffuse dust component in 
	the ISM. The red regions indicate the dust component 
	associated with the short-lived stellar birth clouds. 	
	The large blue stars show the young, massive stars which mostly are found 
	in the birth clouds. The small red stars show the less massive stars (both young and 
	old), which are found both within the birth clouds and elsewhere. 
	For galaxies with higher specific SFRs, we expect the continuum light to be dominated 
	by the young, massive stars in the birth clouds, so both the continuum and emission 
	lines are attenuated by both dust components. Galaxies with lower specific SFRs 
	would have a higher contribution to the continuum emission from less massive stars,  
	which generally reside outside the birth clouds and are only attenuated by the diffuse 
	dust component, while the emission lines are still attenuated by both dust components. 
	Thus this leads to larger differences between \Avstel and \Avemis.
	 }
	\label{fig:illustration}
	
	\end{center}
\end{figure}

For galaxies with the highest SSFRs, the continuum light is dominated 
by young, massive stars. These massive stars would predominately still reside in the birth clouds. 
So for the two-component model, 
both the emission lines and the continuum features would be attenuated by both the 
birth cloud and the diffuse dust components, resulting in $\Avemis \approx \Avstel$. 
Galaxies with lower SSFRs would have a smaller continuum contribution from massive stars, 
so more of the continuum light would come from stars only attenuated by the diffuse dust, 
resulting in \Avemis greater than \Avstel. 
These different cases are illustrated in Figure \ref{fig:illustration}.

We compare our results with those of \citet{Wild11} for local galaxies. 
They also find increasing amounts of extra attenuation with decreasing SSFRs, 
in agreement with the trend we observe. \citeauthor{Wild11} find higher amounts of extra 
attenuation than we do, but the average SSFRs of their sample are lower than for our sample.
The two-component dust model naturally explains this difference, based on the dependence of 
extra attenuation on SSFR as discussed above. 
This model may also explain the higher amount of extra attenuation found by 
\citet{Calzetti00} if their sample has higher average SSFRs than our sample. 

The two-component dust model could also explain the discrepancies found 
between different high-redshift studies, if the samples consist of galaxies with different 
ranges in specific SFR. For example, the \citet{Kashino13} SFR-M$_*$ relation implies a 
higher average SSFR than our sample, and they find a lower amount of extra attenuation. 
\citet{Erb06a} find evidence for no extra attenuation, but the average SSFR is higher 
than for the \citeauthor{Kashino13} sample.

Our explanation for the trend between extra \Av and SSFR was previously mentioned by \citet{Wild11}. They suggest that the trend may alternatively be caused by a decline in 
diffuse dust with decreasing SSFR. However, we observe that \Avstel increases slightly 
with decreasing SSFR, so we do not expect a decline in diffuse dust with decreasing SSFR.

In absolute terms, we find that \Avemis increases with mass and SFR, and decreases with SSFR. 
As stellar mass and SFR are correlated, the trend of increasing dustiness with SFR and mass 
could share the same cause. 
The trend of increasing \Avstel and \Avemis with decreasing SSFR could also share 
the same cause, as the SSFR decreases slightly with increasing mass both in 
the local universe \citep{Brinchmann04} and at higher redshifts 
\citep{Elbaz07, Noeske07, Zheng07, Damen09, Whitaker12}. 
As the trends of \Avstel and \Avemis with increasing mass are the strongest, it is likely that 
mass is the key property. 
This finding may be explained by the fact that more massive star-forming galaxies have 
higher gas-phase metallicities \citep{Tremonti04, Erb06a}.

We do note that we assume a fixed attenuation law in our work. 
Recent work \citep[e.g.,][]{Wild11, Buat12, Kriek13} shows 
that the dust attenuation curve varies with SSFR. As the origin of these observed trends are 
not well understood, and this variation may be the consequence of age-dependent 
extinction in a two-component dust model, we have decided to use the same dust law 
for the derivation of the two dust measurements. However, we cannot rule out the possibility 
that variations in the dust attenuation law may have impacted the trends found in this work.

\subsection{Dust attenuation vs. axial ratio}
\label{sec:axial_ratio}

The two-component dust model also predicts a dependence of dust attenuation properties 
on the axial ratio. Under this model, the amount of dust attenuation from the 
stellar birth clouds is similar in face-on or edge-on systems, while the longer path length 
in edge-on systems results in a larger overall \Avstel and a smaller amount of 
extra attenuation towards the star-forming regions. \citet{Wild11} find evidence for the 
two-component model from the trends of attenuation with axial ratio for objects in the local universe. 

The spatially resolved WFC3 images yield excellent axial ratio measurements.  
However, the axial ratio distribution for our sample is heavily biased towards face-on systems. 
The more edge-on systems may be dustier, so our selection criteria likely introduce this bias 
against edge-on systems. It might also be that our sample objects are not disk-like galaxies. 
Because of sample incompleteness and the small range in axial ratios, 
we are unable to test the two-component model using the inclinations of the galaxies.

\begin{figure}
	\begin{center}

	\includegraphics[width=0.48\textwidth]{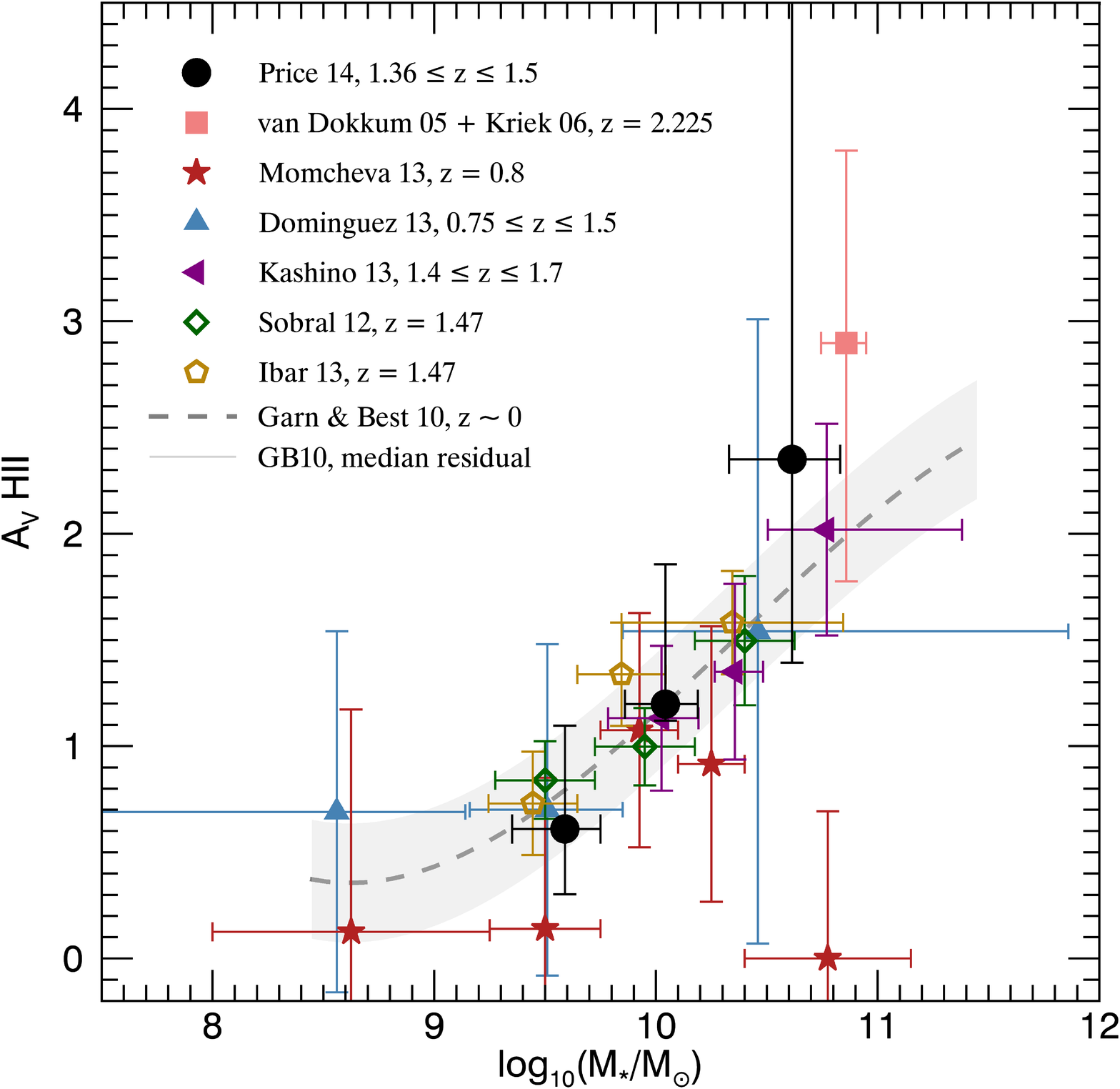}
	\caption{ \Avemis vs. stellar mass comparison between this work and past studies. 
	The filled data 
	\protect\citep[this work;][]{vanDokkum05, Momcheva13, Dominguez13, Kashino13} 
	indicate direct measurements of \Avemis using Balmer decrements, while the open data 
	indicate \Avemis measured with a different method that is calibrated using 
	Balmer decrements. The dashed line gives the median relation derived by \protect\citet{GB10} 
	using SDSS star-forming galaxies. With the exception of the single object 
	from \protect\citet{vanDokkum05} \protect\citep[combined with information from][]{Kriek06}, 
	all other data 
	are the result of stacks (using various stacking schemes, both mean and median) or are mean 
	\protect\citep[combined with stacking]{Momcheva13} or median values 
	\protect\citep{GB10, Sobral12} from samples. }
	\label{fig:avemis_mass}
	
	\end{center}
\end{figure}

\subsection{Comparison of results for \Avemis vs. stellar mass}
\label{sec:comparison}

A number of past studies have measured Balmer decrements 
(and often \Avemis) versus stellar mass \citep{Kashino13, Dominguez13, Momcheva13, 
vanDokkum05}. Other studies have employed different methods of 
measuring \Avemis versus stellar mass. These methods include comparing 
$L_{\unit{IR}}$ to $L_{\unit{H\alpha}}$ as by \citealt{Ibar13}, 
or calibrating [OII]/\Halpha ratio as an alternate 
to Balmer decrements as by \citealt{Sobral12}.
We compare our results with their findings in Figure \ref{fig:avemis_mass}. 
As necessary, we convert stellar masses derived using a Salpeter or 
Kroupa IMF to match our assumption of a Chabrier IMF using the relations given in Equations 
12 and 13 of \citet{Longhetti09}.
 
The collection of results over a range of redshifts makes 
it tempting to speculate that there is no redshift evolution in the mass-\Avemis relation. 
However, the current data are not conclusive. First, there is incompleteness at all masses. 
Second, the measurements of \Avemis are not necessarily equivalent.

The samples of \citet{Dominguez13} and \citet{Kashino13} are similar to ours in redshift 
and method (stacking) for measuring \Avemis. 
Concerning the work by \citet{Dominguez13}, we first note that their sample 
contains fewer objects than our sample. Second, we have a broad range of deep photometric 
coverage for our objects, which reduces the errors in our masses taken from SED modeling. 
We also do not combine measurements between different grisms, to avoid possible normalization 
mismatches affecting line fluxes. 
Compared with \citet{Kashino13}, our sample is larger and on average has slightly lower SSFRs.  
The trend we observe between \Avemis and SSFR shows that the average \Avemis decreases 
with increasing SSFR, which explains why their values of \Avemis are lower 
(though consistent within the errors) than what we measure for similar masses.

\begin{figure}
	\begin{center}

	\includegraphics[width=0.48\textwidth]{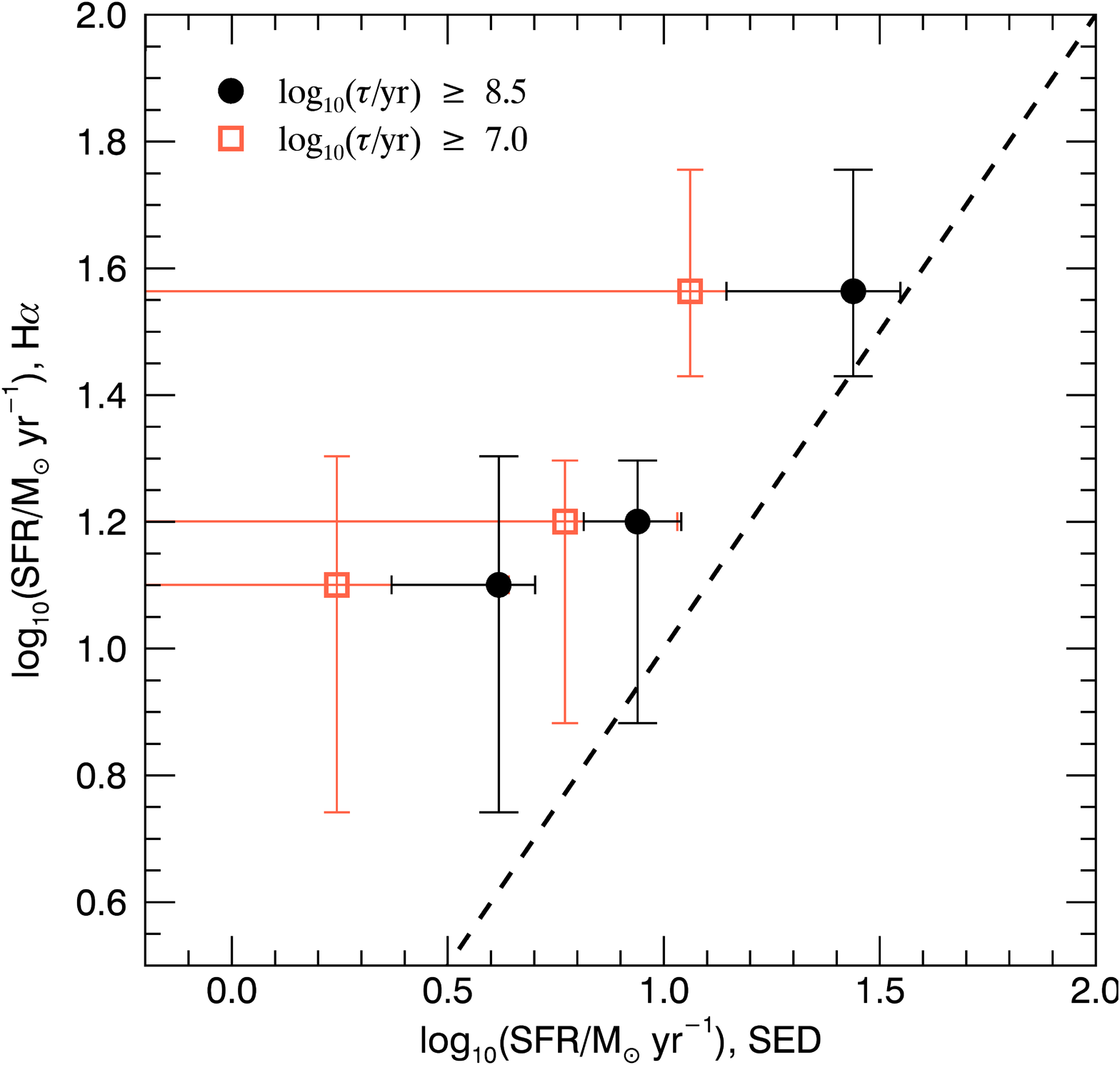}
	\caption{ Comparison between measured \Halpha SFRs \protect\citep{Kennicutt98} 
	and the SED SFRs. 
	The two sets of SED SFRs differ only in the choice of the minimum star formation 
	\textit{e}-folding timescale, $\tau$:  $\logtau \geq 7.0$ (open red squares) and 
	$\logtau \geq 8.5$ (closed black circles). 
	(The dashed black line shows equal \Halpha SFRs and SED SFRs.) 
	The \Halpha SFRs agree better with the SED SFRs 
	calculated with the higher $\tau_{min}$, which is similar to previous findings 
	that setting $\tau_{min} \sim 300 \ \unit{Myr}$ yields the most reasonable SED fits for 
	star-forming galaxies.
	(See Section \protect\ref{sec:SFRHa_SFRSED} for a full discussion.) 
	The errors in log SED SFR are the $1 \sigma$ scatter within the bins.
	}
	\label{fig:sfr_comp}
	
	\end{center}
\end{figure}

\subsection{Implications for H$\alpha$ SFR compared with SED SFR}
\label{sec:SFRHa_SFRSED}

With our measurements of dust attenuation  towards \textsc{Hii} regions, 
we can for the first time 
directly measure \Halpha SFRs for a large sample of $z \sim 1.5$ galaxies. 
We calculate the \Halpha SFRs using the relation from \citet{Kennicutt98} converted 
to a \citet{Chabrier03} IMF. 
\Halpha fluxes are taken from the stacks in SED SFR, and are 
converted to physical units by comparing the flux of the stack with the 
weighted average (see Section \ref{sec:line_measurement}) 
of the individual objects' \Halpha fluxes, each corrected for [NII].
The luminosity distance is taken as the weighted average of the individual objects' 
luminosity distances.  We then use the \Halpha SFRs to test the much more debated SED SFRs.

The comparison of SFR indicators is shown in Figure \ref{fig:sfr_comp}. 
The \Halpha SFRs are compared to the weighted average of SED SFRs derived assuming an 
exponentially decaying SFH and a minimum \emph{e}-folding time of $\logtaumin = 8.5$ 
(shown as filled black circles). For these parameters, the SED SFR values underestimate 
the \Halpha SFRs, and are inconsistent at the $\sim 1.5 \sigma$ level. 
However, if a shorter decay time of $\logtaumin = 7$ (open red squares) is adopted, 
the SED SFRs are low and inconsistent with the \Halpha SFRs at the 
$\sim 4 \sigma$ level. 
Thus for our sample, SED SFRs derived with $\logtaumin = 8.5$ are more in line 
with the \Halpha SFRs.

This is in general agreement with the work of \citet{Wuyts11a}. 
They compare UV+IR and SED SFRs and find good agreement between the 
SFR indicators for $\logtaumin = 8.5$, but when a shorter time of $\logtaumin = 7$ is adopted, 
they find that the SED SFRs underestimate the true SFRs. 
\citet{Reddy12} also compare UV+IR SFRs with SED SFRs. However, instead of using a 
longer \taumin for a decreasing SFH, they find the SED SFRs agree with UV+IR SFRs 
when increasing SFHs are adopted.

The SFR indicators may also differ because they probe different 
stellar mass ranges, and thus different star formation timescales.  The \Halpha SFRs depend 
on OB stars, so are averaged over $\sim 10 \ \unit{Myr}$, while SED SFRs are limited by the 
discrete nature of SED fitting and depend on the UV flux from O, B and A stars, which live longer. 
These timescale differences could cause discrepancies for galaxies with episodic or 
rapidly increasing or decreasing SFHs. 
However, as we stack multiple objects and thus average over many SFHs, we expect that 
the timescale differences should not significantly influence the measured SFRs.

\begin{figure*}
	\begin{center}
	
	\includegraphics[width=0.99\textwidth]{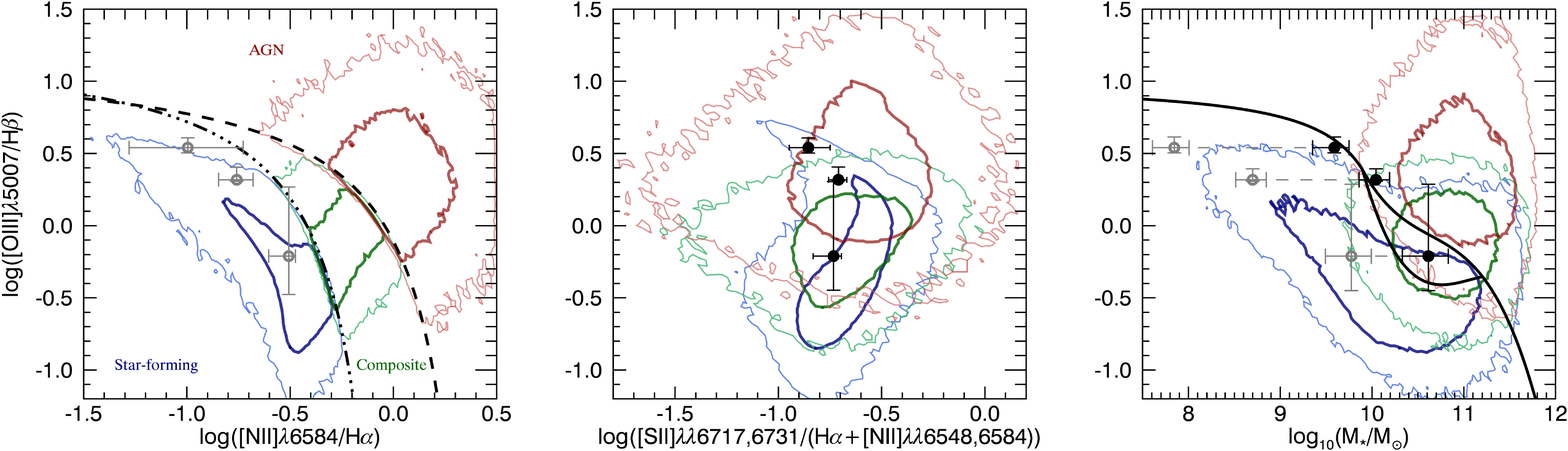}
	\caption{
	Left: 
	Traditional BPT \protect\citep{Baldwin81} diagram used to identify the ionizing mechanism for 
	optical emission lines. The dashed line is the theoretical limit between star-forming (SF)  
	galaxies and AGN, as derived in \protect\citet{Kewley01}, and the dashed-dot line 
	is the empirical division derived in \protect\citet{Kauffmann03} for galaxies in the SDSS. 
	The color contours represent SDSS galaxies, divided into three regions according 
	to these dividing lines. The darker, thicker lines enclose 68\% of the population, 
	while the lighter, thinner lines enclose 95\%. The grey open points represent our 
	stacked spectra (in bins of $M_*$), with the [NII]/\Halpha measurement inferred 
	from the average stellar masses (i.e. not directly measured from the data.) 
	Middle: 
	alternative BPT diagram, using the blended lines observed with the grism. Color 
	contours correspond to the SDSS galaxies in the left panel. The black circles show 
	the measurements from our stacked spectra. This panel illustrates that the 
	combination of blended emission lines does not enable us to discriminate between 
	SF galaxies and AGN. 
	Right: 
	the Mass-Excitation (MEx) diagnostic from \protect\citet{Juneau11}, where 
	the solid lines are the empirical divisions between SF galaxies and AGN, valid 
	for $z<1$ galaxies. Our stacked spectra (black circles) lie on the dividing line between 
	the SF and AGN regions of the diagram. 
	Recently \citet{Newman14} have shown that high redshift SF galaxies tend to be offset 
	to the right from lower redshift SF galaxies in this diagram.
	We also apply the empirical correction they derive
	for $z \sim 2$ galaxies (open grey circles) to better compare our sample 
	to the $z \sim 0$ MEx divisions. 
	(See Section \protect\ref{sec:agn_contamination} for more discussion.)
	}
	\label{fig:bpt}
	
	\end{center}
\end{figure*}

\subsection{AGN contamination}
\label{sec:agn_contamination}

As discussed in Section \ref{sec:sample_selection}, we use a number of methods to reject AGNs 
from our sample. However, as almost all individual objects do not have sufficient 
line SNRs, and more importantly we do not have separate [NII] and \Halpha measurements, we cannot use a BPT \citep{Baldwin81} diagram to distinguish whether the line emission 
originates from star formation or AGN.

After stacking, there is sufficient line SNR to place the binned values in the BPT and alternative 
diagrams, which we show in Figure \ref{fig:bpt}. 
These diagrams allow us to assess if there is AGN contamination in the stacks. 
We use the data from the stacks in stellar mass  in these diagrams. 

In the left panel, we show the traditional BPT diagram: [NII]/\Halpha vs. [OIII]/\Hbeta. 
Included are the theoretical limit for star-forming (SF) galaxies from \citet{Kewley01} (dashed line) 
and the more conservative empirical division between SF galaxies and AGN by \citet{Kauffmann03} 
(dashed-dot line), both of which are derived for galaxies at $z \sim 0$. 
The SDSS DR7 galaxies \citep{Abazajian09} are divided into three categories 
based on these dividing lines: SF galaxies (blue contours), composite (green contours), 
and AGN (red contours).  These same group definitions are used in the alternate 
BPT diagrams (middle and right of Figure \ref{fig:bpt}). 
Because we do not measure [NII] directly (instead inferring a value as described in 
Section \ref{sec:nii_correction}), we cannot use the traditional BPT diagram as a post-analysis 
check. 
Instead, we show that our inferred [NII]/\Halpha ratios and the observed 
[OIII]/\Hbeta ratios, shown as open grey circles, are consistent with values for SF galaxies. 
However, in agreement with other observational \citep[e.g.,][]{Shapley05,Liu08} 
and theoretical studies \citep{Kewley13} of distant galaxies, the data points are moved 
slightly up (and possibly to the right) in the BPT diagram 
compared to the average values found in the local star-forming sequence. 

An alternate BPT diagram is [SII]/\Halpha vs. [OIII]/\Hbeta. 
However, \Halpha is not a directly measured quantity, so using the deblended \Halpha 
values makes it impossible to detangle the assumptions of the [NII] correction with 
the possible presence of AGN. Instead, we make an alternate diagram with only directly measured 
quantities: [SII]/(H$\alpha$+[NII]) vs. [OIII]/\Hbeta, which we show in the middle panel of 
Figure \ref{fig:bpt}. There is significant overlap between the different categories of SDSS galaxies, 
so we are unable to discriminate between SF galaxies and AGN using combinations 
of blended emission lines.

A final alternate to the BPT diagram, the Mass-Excitation (MEx) diagram \citep{Juneau11}, 
is useful as we directly measure the necessary lines ([OIII] and \Hbeta) and derive 
stellar mass with SED fitting, which we show in the right panel of Figure \ref{fig:bpt}.  
The solid black lines are their empirical divisions between SF, composite, and AGN, which 
are valid up to $z \sim 1$.  We include the SDSS categories, using stellar masses from the 
MPA/JHU value-added catalogs \citep{Kauffmann03}, to demonstrate that while the 
separation is not as clean as the traditional BPT diagram, it does separate SF 
galaxies from AGN in the local universe. 
Our data are consistent with the SF region as defined by \citet{Juneau11}, 
though the data do lie on the division boundary. 
However, work by \citet{Newman14} at $z \sim 2$ has shown that the low redshift empirical 
divisions in the MEx diagram incorrectly classify $z \sim 2$ SF galaxies as AGN. 
They use the shift in the mass-metallicity relation between $z \sim 0$ and 2 to derive 
a shift to lower mass to bring the high-$z$ values into agreement with the 
$z \sim 0$ MEx diagram \citep[see Fig. 14 of][]{Newman14}. 
This empirical correction (shown as open grey circles in the 
right panel of Figure \ref{fig:bpt}) shifts our data securely into the SF region of the diagram. 
These diagnostics demonstrate that there is likely not much AGN contamination 
to the line emission in our sample.

\subsection{[NII] contamination}
\label{sec:nii_contamination}

One of the largest sources of uncertainty in this analysis is our correction for 
[NII] in our measurement of \Halpha from the blended H$\alpha$+[NII] line. 
Figure \ref{fig:avplot} demonstrates how much our results change when we do not correct for [NII]. 
We use the relation between [NII]/\Halpha and stellar mass by \citet{Erb06a} 
as a way of detangling the blended lines. We infer the [NII] contribution from the weighted 
average mass of each stack. Because the two samples have similar masses 
and SFRs, and are close in redshift, the average sample properties should be 
fairly similar. Still, the use of this relation may introduce bias into our results. 

However, our gas phase metallicities may be different from those of \citet{Erb06a}, 
due to slight differences in redshift, mass or (S)SFR, which could affect both the slope and 
scaling of the [NII]/\Halpha relation. 
Such trends have been proposed by \citet{Mannucci10}, in the form of the fundamental 
metallicity relation (FMR). The FMR suggests that metallicity depends on both stellar mass 
and SFR.

We repeat our analysis using the FMR in conjunction with the \citet{Maiolino08} 
[NII]/\Halpha metallicity calibration as an alternative way of calculating the [NII] correction. 
This allows us to test whether the trend of \Avextra versus SSFR may be due to 
a variation in [NII]/\Halpha with SFR.
If the FMR is adopted, the \Avemis values are slightly lower than the \Avemis values 
derived using the \citet{Erb06a} values, which reduces the strength of the trend of \Avextra 
with SSFR. In this case, the relation we observe between \Avextra and SSFR is weaker 
than the trend presented in Section \ref{sec:ssfrsfrmass}, and is consistent with 
no trend, as the difference is only $\sigma = 0.4$. 
However, the data are still suggestive of a possible decreasing trend of \Avextra with SSFR, 
which would imply the signal is not entirely caused by trends with metallicity.

It is also possible that there is some AGN contamination at the highest masses 
\citep{Kriek07}, which would also result in an underestimate of the [NII] fraction. 
In most cases, our corrected \Halpha flux would be higher than the true value, leading 
to an overestimate of \Avemis.

We use the mass-[NII]/\Halpha relation presented in \citet{Erb06a} for our analysis 
over the FMR due to uncertainty about metallicity relations at high redshifts. 
First, the FMR is still not well tested at these redshifts \citep{Cullen14}. 
Second, recent work has also questioned whether [NII]/\Halpha correlates with metallicity 
at these redshifts \citep{Kulas13, Newman14}. \citet{Erb06a} present direct measurements 
of [NII]/\Halpha, which allows us to avoid systematic problems with metallicity calibrations.

\subsection{Incompleteness and other systematic uncertainties}
\label{sec:incompleteness_bias}

One of the strengths of our analysis is that we draw a sample from a non-targeted grism survey, 
with sample cuts designed to avoid bias as much as possible. However, bias and incompleteness 
most likely still affect our sample.

The dustiest star-forming galaxies have very attenuated \Halpha fluxes. 
Our \Halpha SNR selection cut, designed to avoid adding noise to our analysis, 
introduces bias against galaxies with large \Avemis. 
This bias affects the high mass, high SFR end of the sample, as these 
objects have the largest \Av. As lower mass, lower SFR galaxies tend to have lower 
\Halpha luminosities, the \Halpha SNR cut will also exclude objects with the 
largest \Av values in that mass range.

Our continuum normalization scheme also introduces bias into our analysis. 
We adopt a normalization scheme to improve the signal of our stack, but 
the cost is that some objects have much higher scaled \Halpha fluxes, and 
thus they dominate our stacks. This biases our results towards those objects with higher 
\Halpha equivalent widths in a given bin. Because the line measurements are biased based 
on the \Halpha flux, we take the weighted average of the continuum values within a bin to 
ensure a fair comparison. 
However, if we instead used a non-weighted average for $\Avstel$, 
the \Avstel vs. \Avemis relationship does not change much.

\section{Summary}
\label{sec:summary}

In this paper, we investigate dust attenuation in $z \sim 1.5$ star-forming galaxies 
using data from the 3D-HST survey. We measure both the dust towards \textsc{Hii} 
regions, using Balmer decrements, and the integrated dust properties, using SED modeling. 
We find that there is extra attenuation towards star-forming regions. On average 
the total attenuation of these regions, \Avemis, is 1.86 times the integrated 
dust attenuation, \Avstel.

However, the amount of extra attenuation is not the same for all galaxies. 
We find that the amount of extra attenuation decreases with increasing SSFR, 
in agreement with the results by \citet{Wild11} for low-redshift galaxies. 
Our findings are consistent with the two-component dust model, which 
assumes there is a diffuse dust component in the ISM and a 
dust component associated with the short-lived stellar birth clouds. 
For galaxies with high SSFR, the stellar light will be dominated by continuum emission 
from the younger stellar population in the birth clouds, resulting in similar attenuation 
toward the line and continuum emission. For more evolved galaxies, much of the 
stellar light will only be attenuated by the diffuse ISM, leading to larger 
discrepancies between the two dust measures. 
The observed trend of \Avextra with SSFR may be affected by uncertainties 
in the [NII] correction and possible dust attenuation law variations. 
Future work is necessary to determine what role these effects have on the relation 
between \Avextra and SSFR.

Similar to previous studies 
\citep{ForsterSchreiber09, Yoshikawa10, Wuyts11a, Wuyts13, Kashino13}, 
we find less extra attenuation in distant galaxies than is found in the local universe 
\citep{Calzetti00}. This effect can also be explained by the two-component model, as lower 
redshift objects tend to have lower SSFRs than higher redshift galaxies 
\citep[e.g.,][]{Noeske07a, Whitaker12b, Fumagalli12}.

We find that both \Avemis and \Avstel increase with increasing SFR and stellar mass, and 
decreasing specific SFRs. However, our data are biased against the dustiest objects, 
which may affect these trends. 
We also observe there to be little redshift evolution in the \Avemis-$M_*$ relation, 
although uncertainties and incompleteness makes it impossible to make a definite claim.

Using the Balmer decrements, we calculate dust-corrected \Halpha SFRs to test 
the accuracy of SFRs derived from SED fitting. We find better agreement between the 
SFR indicators if short SFH decay times are not allowed and the constraint 
$\log_{10}(\tau/\unit{yr}) \ge 8.5$ is used. This generally agrees with the results 
of past studies comparing UV+IR and SED SFRs \citep{Wuyts11a} or \Halpha and 
SED SFRs \citep{Wuyts13}. However, even with this constraint the SED SFRs slightly 
underestimate the \Halpha SFRs.

We note that our results are slightly impacted by incompleteness and systematic uncertainties. 
First, we employ SNR cuts on \Halpha to ensure quality data, which likely 
results in incompleteness of the dustiest galaxies. 
Second, to obtain significant \Hbeta detections, we stack spectra, and thus our 
normalization scheme or incorrectly measured integrated properties 
could impact our measurements. 

Most of these issues can be overcome with future observations by a number of new 
multi-object near infrared spectrographs on 8-10 m class telescopes, among which is 
MOSFIRE on Keck \citep{McLean10}. 
These instruments have higher spectral resolutions, which will avoid blended lines. 
They will also allow for deeper measurements, which will yield more accurate Balmer decrements 
of individual objects as well as allowing for investigation of dust to higher \Av limits.
Additionally, including rest-frame mid- and far-IR data in future work will ensure more 
accurate values of \Avstel. Better measurements of \Avemis and \Avstel are necessary 
to better constrain the geometric distribution of dust in high redshift galaxies and 
the effects of the dust-to-star geometry on dust attenuation.

\acknowledgements

We thank Edward Taylor and Nick Hand for useful discussions, 
and David Sobral and Daichi Kashino for sharing data for comparison. 
We thank the anonymous referee for constructive comments, which 
have improved this paper. 
This work is based on observations taken by the 3D-HST Treasury Program 
(GO 12177 and 12328) with the NASA/ESA HST, 
which is operated by the Association of Universities for Research in Astronomy, Inc., 
under NASA contract NAS5-26555. 
We acknowledge support from STScI grant 12117.21-A. 
SP is funded by the National Science Foundation Graduate Research Fellowship 
under Grant No. DGE 1106400.





\appendix

\begin{figure*}[h!]
	\begin{center}

	\includegraphics[width=0.92\textwidth]{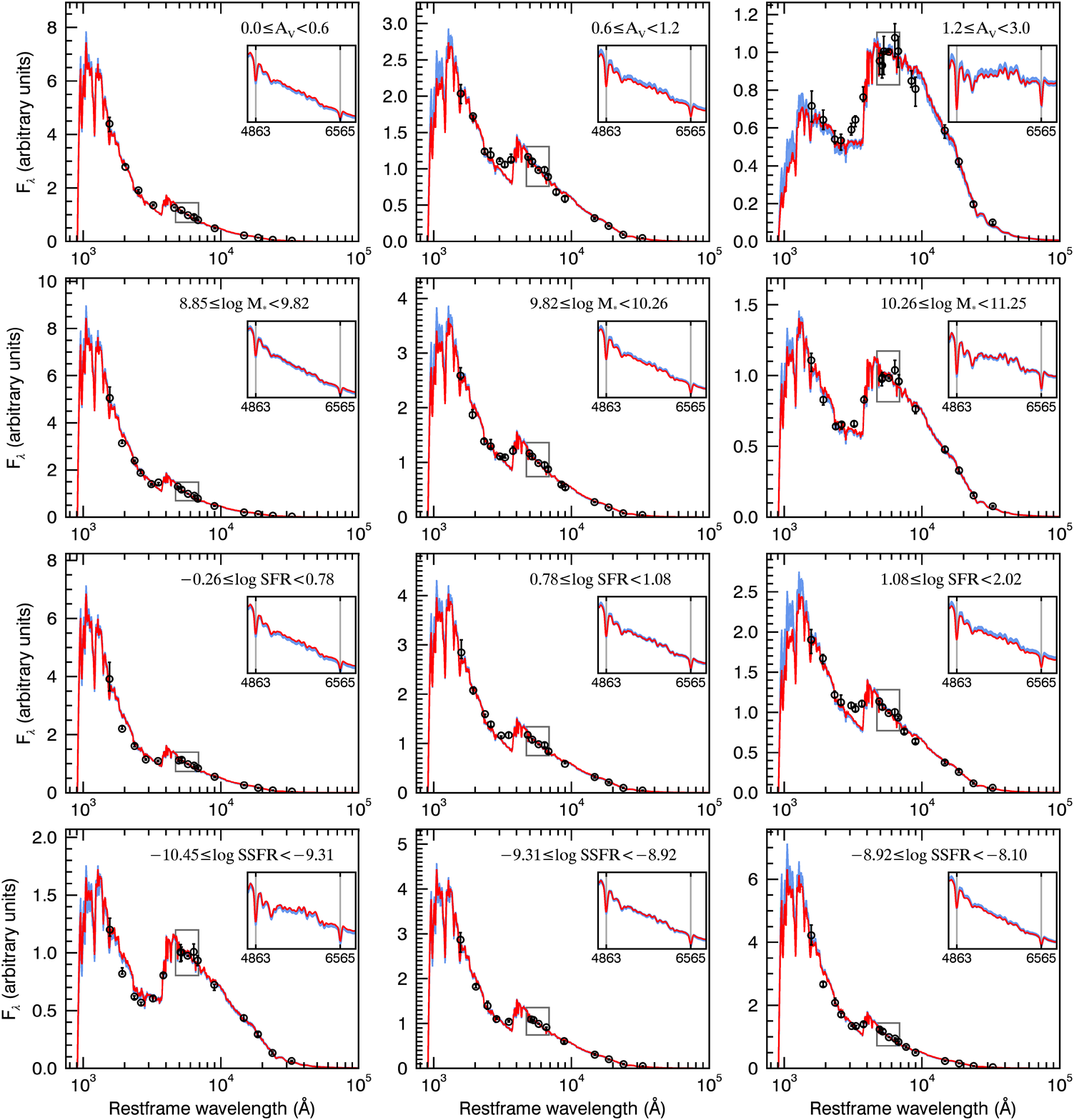}
	\caption{
	Binned photometry and stacked best-fit stellar models for bins in 
	stellar $\Av\!$ (top), $\log M_{*}$ (second row), 
	log SFR (third row) and log SSFR (bottom). 
	The photometry is averaged in wavelength bins, and the error is taken to be the error on the 
	mean. The binned photometry is shown by the black circles. 
	The stacked best-fit FAST models are plotted in red. The $3\sigma$ errors in 
	the stacked models are plotted in blue. The insets show the best-fit models 
	(and $3\sigma$ errors) plotted linearly with wavelength near \Hbeta and \Halpha 
	(denoted by left and right vertical grey lines, respectively).
	}
	\label{fig:phot}
	
	\end{center}
\end{figure*}

In addition to the stacked spectra (presented in Fig. \ref{fig:spec}), we also present 
the average photometry for each bin (see Fig. \ref{fig:phot}). 
The photometry covers a much greater wavelength range, so the average photometry allows 
us to examine the average SED shape outside the limited range covered by the grism spectra. 
The photometry for each object is first normalized to match the grism normalization 
(see Section \ref{sec:stacking}), and the central wavelength of each filter is de-redshifted using 
the object's grism redshift. The data are averaged within each bin, using the normalized 
fluxes and rest-frame wavelengths. 
If not all objects in the bin have coverage in a given filter, nearby filters are combined so each 
averaged photometric point contains at least the same number of measurements as number of 
objects in the bin, provided the photometry are not too widely separated in wavelength.

For comparison with the average photometry, we also show the average 
best-fit continuum model and the errors on the best-fit model. Plotting the continuum 
model and error allows us to examine the error in the amount 
of Balmer absorption. The average best-fit stellar population model and simulated best-fit 
models are calculated as described in Section \ref{sec:stacking}. 
The $\pm 3 \sigma$ continuum errors are estimated at every wavelength using the 
simulated continuum models.

The perturbations of the photometry of the individual objects do not lead to large 
variation in the stacked best-fit models, especially near the Balmer lines. The broad wavelength 
coverage of the photometry, specifically across the Balmer break, provides reasonably 
tight constraints on the average amount of Balmer absorption for the stacks.

\end{document}